\documentclass[pre,aps]{revtex4}\usepackage{amsfonts}
\usepackage{times}
\usepackage{amsmath}
\usepackage{graphicx}
\usepackage{amssymb}
\usepackage{xcolor}

\begin{document}

\title{Solitary waves in the Ablowitz--Ladik equation with power-law
nonlinearity}
\author{J. Cuevas-Maraver}
\affiliation{Grupo de F\'{\i}sica No Lineal, Departamento de F\'{\i}sica Aplicada I,
Universidad de Sevilla. Escuela Polit\'{e}cnica Superior, C/ Virgen de \'{A}%
frica, 7, 41011-Sevilla, Spain \\
Instituto de Matem\'{a}ticas de la Universidad de Sevilla (IMUS). Edificio
Celestino Mutis. Avda. Reina Mercedes s/n, 41012-Sevilla, Spain}
\author{P. G. Kevrekidis}
\affiliation{Department of Mathematics and Statistics, University of Massachusetts
Amherst, Amherst, MA 01003-4515, USA}
\author{Boris A. Malomed}
\affiliation{Department of Physical Electronics, School of Electrical Engineering,
Faculty of Engineering, and Center for Light-Matter Interaction, Tel Aviv
University, P.O.B. 39040, Tel Aviv 69978, Israel}
\author{Lijuan Guo}
\affiliation{Department of Mathematics and Statistics, University of Massachusetts
Amherst, Amherst, MA 01003-4515, USA}
\affiliation{School of Mathematical Sciences, USTC, Hefei, Anhui 230026, P.R. China}

\begin{abstract}
We introduce a generalized version of the Ablowitz-Ladik model with a
power-law nonlinearity, as a discretization of the continuum nonlinear Schr{%
\"{o}}dinger equation with the same type of the nonlinearity. The model
opens a way to study the interplay of discreteness and nonlinearity features.
We identify stationary discrete-soliton states for different values of
nonlinearity power $\sigma $, and address changes of their stability as
frequency $\omega $ of the standing wave varies for given $\sigma $. Along
with numerical methods, a variational approximation is used to predict the
form of the discrete solitons, their stability changes, and bistability
features by means of the Vakhitov-Kolokolov criterion (developed from the
first principles). Development of instabilities and the resulting asymptotic
dynamics are explored by means of direct simulations.
\end{abstract}

\maketitle

\section{Introduction}

The study of models of the discrete nonlinear-Schr{\"{o}}dinger (DNLS) type
has been a focal point within the broader theme of nonlinear dynamical
lattices over the past two decades~\cite{book}. First, DNLS systems are
obviously relevant to the discretization of the ubiquitous continuum
nonlinear Schr{\"{o}}dinger equations~\cite{sulem,ablowitz,ussiam}. Still
more important is, arguably, the relevance of DNLS equations in their own
right as physical models in a multitude of settings. In particular, they
have been extensively used for modeling light propagation in arrays of
optical waveguides~\cite{dnc,moti}. They have been also greatly employed for
the development of the broad topic of dynamics of atomic Bose-Einstein
condensates (BECs) trapped in deep optical lattices~\cite{ober}. Other
important applications of DNLS models are the study of the denaturation of
the \ DNA double strand \cite{Peybi}, breathers in granular crystals~\cite%
{chong}, the dynamics of protein loops~\cite{niemi}, etc.

A model that plays a critical role in the understanding of DNLS dynamical
lattices is their integrable sibling, namely, the Ablowitz-Ladik (AL)
equation~\cite{ablolad} (see also~\cite{ablowitz}). In addition to
offering the unique integrability structure~\cite{ablowitz}, the AL model is
a reference point for the examination of numerous features of nonlinear
discrete settings, such as stability of solitary waves~\cite{kapitula},
Bloch oscillations~\cite{bishop}, etc. The AL equation is also important as
an ingredient of the Salerno model, which combines it with the nonintegrable
DNLS equation~\cite{salerno}. The Salerno model was widely used, as
a platform for developing
perturbation theory and exploring direct numerical simulations~\cite{cai}, in various
contexts, including, in particular, the challenging problem of collisions
between solitary waves~in dynamical lattices \cite{dmitriev}. Further, the
AL equation by itself, and in the form of its Salerno generalization,
provides for the mean-field limit of the class of quantum systems in the
form of the so-called \textquotedblleft nonstandard" Bose-Hubbard lattices
(BHLs), in which the inter-site coupling depends on the sites' populations
(hence the coupling is nonlinear, in terms of the mean-field approximation).
Numerous realizations of such BHLs, including their mean-field realizations,
were reviewed in a recent survey article \cite{review}. Actually, the list
of applications of the AL model is anything but exhaustive, as relevant
applications continue to emerge through the examination, e.g. more recently,
of features such as rogue waves in the AL model~\cite{akhm,yang}, and their
comparison to the DNLS case~\cite{hoffman}.

In the present work, we consider a variant of the AL model with a power-law
nonlinearity instead of the particular case of the cubic one. This is
motivated by many previous studies ---see e.g.~\cite{sulem} and
references therein for the continuum case, as well as Refs.~
\cite{Turitsyn,malomed,weinstein,ourDNLS,Kladko} for the
discrete case--- of the generalized NLS model with nonlinear term $|\psi|^{2\sigma }\psi $ instead of the usual cubic one corresponding to the
particular value of $\sigma =1$. In particular, in the one-dimensional (1D)
case the Lee-Huang-Yang correction to the mean-field nonlinearity in
two-component Bose-Einstein condensates (BECs) corresponds to the
self-attractive term with $\sigma =0.5$ \cite{Grisha1,Grisha2}. The
model with a general nonlinear term is particularly appealing because it offers, at the continuum level
and for one spatial dimension, the potential to study the transition from
integrability (at $\sigma =1$) and the existence of stable solitary waves
(at $\sigma <2$) to the emergence of catastrophic self-focusing
(at $\sigma \geq 2$). In the DNLS variant of the model, there exist
interesting features too, such as bistability (for a range of values of $%
\sigma $), as well as the emergence of energy thresholds for the existence
of discrete solitons, again for $\sigma >2$ in one spatial dimension.
However, the AL generalization to a power-law nonlinearity was not explored
before, to the best of our knowledge. It is the aim of the present work to
report this analysis.

More specifically, our aim is to examine the nature of discrete solitons for
the AL-type lattice nonlinearity, $\propto |\psi _{n}|^{2\sigma }(\psi
_{n+1}+\psi _{n-1})$, with different values of $\sigma $ (characterizing also the structure
of these discrete solitons by means of a variational approach) and to explore
whether these modes change their stability with the variation of their
intrinsic frequency (alias chemical potential, in terms of BEC) for given $%
\sigma $. In particular, we aim to find out if the stability change can be
predicted by a variant of the Vakhitov-Kolokolov (VK) criterion, which is
well known in other contexts \cite{Vakh,book} --- In the DNLS (and NLS)
settings, the VK criterion suggests that a change of the monotonicity of the
$l^{2}$ (or $L^{2}$, respectively) norm vs. the soliton's intrinsic
frequency dependence induces a change of its stability \cite{book}.Finally, we
explore the full stability (via the spectral stability analysis) and
evolution of instabilities in the model, by means of numerical computations.
This allows us, among other features, to corroborate predictions of the VK criterion and analyze
manifestations of the bistability in the model.

Our presentation is structured as follows. In section II, we offer our
analytical considerations, including the model setup, the present variant of
the VK criterion, and the variational approximation. In section III, we show
numerical computations concerning both the existence and spectral stability
of discrete solitons and their nonlinear evolution. Finally, in section IV
we summarize our findings and discuss directions for further work.

\section{Analytical Considerations}

\subsection{The general Setup, conservation laws, and stability analysis}

Our generalized AL model with the power-law nonlinearity is introduced as
\begin{equation}
\mathrm{i}\dot{\psi}_{n}+C(\psi _{n+1}+\psi _{n-1})+\beta (\psi _{n+1}+\psi
_{n-1})|\psi _{n}|^{2\sigma }=0.  \label{eq:dyn}
\end{equation}%
Here, $C$ is the (linear)
coupling constant between adjacent sites, while $\sigma $
represents, as defined above, the nonlinearity power. In what follows below,
we fix $\beta =1$ by means of obvious rescaling, considering the focusing
sign of the nonlinearity ($\beta >0$). This model can be thought of as a
discretization of the NLS model with the general power-law nonlinearity~\cite%
{sulem}. At the same time, as mentioned above, it is an intriguing setup for
exploring differences (for sufficiently large $\sigma $, such as $\sigma >2$
in the 1D case considered here) of the collapse phenomenology in discrete
vs. continuum models. With these motivations in mind, we will construct
localized states of the model, investigate their stability through
computation of eigenvalues of small perturbations, and numerically examine
evolution of unstable states.

An important property of Eq.~(\ref{eq:dyn}) is that it preserves a suitably
defined norm-like quantity. The derivation of this property, that we will develop here, is
generally applicable to nonlinear lattice equations of the form
\begin{equation}
\mathrm{i}\dot{\psi}_{n}+(C+f(|\psi _{n}|^{2}))(\psi _{n+1}+\psi _{n-1})=0,
\label{eq:dyn1}
\end{equation}%
where $f$ may be an arbitrary real function of $|\psi _{n}|^{2}$. We
assuming that $P$ (which should revert to the usual $L^{2}$ norm in the
continuum limit) depends only on the on-site intensity $|\psi _{n}|^{2}$:
\begin{equation}
P=\sum_{n}g(|\psi _{n}|^{2}).
\end{equation}%
From here it follows that
\begin{equation}
\mathrm{i}\frac{dP}{dt}=\sum_{n}g^{\prime }(C+f)\left[ \left( \psi _{n+1}^{\star
}\psi _{n}-\psi _{n+1}\psi _{n}^{\star }\right) -\left( \psi _{n}^{\star
}\psi _{n-1}-\psi _{n}\psi _{n-1}^{\star }\right) \right] ,
\end{equation}%
where Eq. (\ref{eq:dyn}) is used to substitute $d\psi _{n}/dt$, and $%
g^{\prime }$ stands for the derivative of $g$. Thus, the selection of $%
g^{\prime }=1/(C+f)$ leads $dP/dt=0$, due to the telescopic nature of the
resulting summation. For the particular case of $f(x)=x^{\sigma }$, this
results in
\begin{equation}
P=\frac{1}{C}\sum_{n}|\psi _{n}|^{2}{}_{2}F_{1}\left( 1,1/\sigma ;1+1/\sigma
;-\frac{|\psi _{n}|^{2\sigma }}{C}\right) \equiv \sum_{n}\mathcal{P}_{n},
\label{eq:norm}
\end{equation}%
where $\mathcal{P}_{n}$ is the  density associated with
the conserved quantity at each node, and $_{2}F_{1}$ is
the Gauss hypergeometric function. In the special case of $\sigma =1$, it is
straightforward to see that this falls back to the simple expression
relevant for the cubic integrable AL model~\cite{cai}, $\left( \mathcal{P}%
_{n}\right) _{\sigma =1}=\ln \left( 1+\frac{1}{C}\left\vert \psi
_{n}\right\vert ^{2}\right) $ (up to a $C$-dependent constant prefactor). We
have thus obtained an important, even if somewhat cumbersome, conservation
law for Eq.~(\ref{eq:dyn}).

A further important observation for Eq.~(\ref{eq:dyn}) is that a natural way
to analyze the system can be to explore its similarity to the integrable case of
$\sigma =1$ and the perturbation theory around it~\cite{cai,review}. In
particular, we conclude that the Hamiltonian of the general AL equation
remains the same as in the integrable case,
\begin{equation}\label{eq:Ham}
H=-\sum_{n}\left( \psi _{n}\psi _{n+1}^{\star }+\psi _{n}^{\star }\psi
_{n+1}\right) ,
\end{equation}%
being a conserved quantity too. It is also relevant to define the Poisson
brackets corresponding to this Hamiltonian:
\begin{equation}
\{B,\tilde{B}\}=\mathrm{i}\sum_{n}\left( \frac{\partial B}{\partial \psi _{n}}\frac{%
\partial \tilde{B}}{\partial \psi _{n}^{\star }}-\frac{\partial B}{\partial
\psi _{n}^{\star }}\frac{\partial \tilde{B}}{\partial \psi _{n}}\right)
(C+f),
\end{equation}%
where $B$ and $\tilde{B}$ are two arbitrary functionals. This definition leads to $%
\{\psi _{n},\psi _{m}^{\star }\}=i(C+f)\delta _{nm}$, while $\{\psi _{n}\psi
_{m}\}=\{\psi _{n}^{\star },\psi _{m}^{\star }\}=0$. The respective
Hamiltonian form of Eq.~(\ref{eq:dyn1}) is
\begin{equation}
\dot{\psi}_{n}=\{H,\psi _{n}\}.
\end{equation}

In our numerical computations, we sought for stationary solutions with
frequency $-\omega $, in the form of $\psi _{n}(t)=\phi _{n}\mathrm{e}^{%
\mathrm{i}(\omega +2C)t}$ with real $\phi _{n}$ obeying the following set of
algebraic equations:

\begin{equation}
-\omega \phi _{n}+C(\phi _{n+1}+\phi _{n-1}-2\phi _{n})+(\phi _{n+1}+\phi
_{n-1})\phi _{n}^{2\sigma }=0.  \label{eq:st}
\end{equation}%
The linear stability was explored by considering perturbations with
infinitesimal amplitude $\delta $ around the stationary solutions. To this
end, we substitute ansatz $\psi _{n}(t)=\mathrm{e}^{i(\omega +2C)t}\left[
\phi _{n}+\delta (a_{n}\mathrm{e}^{\lambda t}+b_{n}^{\ast }\mathrm{e}%
^{\lambda ^{\ast }t})\right] $ in Eq.~(\ref{eq:st}), and then solve the
ensuing linearized eigenvalue problem: $\lambda (a_{n},b_{n})^{T}=\mathcal{M}%
(a_{n},b_{n})^{T}$, with matrix
\begin{equation}
\mathcal{M}=\mathrm{i}\left(
\begin{array}{cc}
L_{1} & L_{2} \\
&  \\
-L_{2}^{\ast } & -L_{1}^{\ast }%
\end{array}%
\right) ,  \label{eq:stab}
\end{equation}%
where submatrices $L_{1}$ and $L_{2}$ in Eq. (\ref{eq:stab}) are composed of
the following entries:

\begin{equation}
(L_{1})_{n,m}=\left[ -\omega -2C+\sigma (\phi _{n+1}+\phi _{n-1})\phi
_{n}^{2\sigma -2}\phi _{n}\right] \delta _{n,m}+(C+\phi _{n}^{2\sigma
})(\delta _{n,m+1}+\delta _{n,m-1}),
\end{equation}%
\begin{equation}
(L_{2})_{n,m}=\sigma (\phi _{n+1}+\phi _{n-1})\phi _{n}^{2\sigma -1}\delta
_{n,m}
\end{equation}%
The stationary solutions are spectrally stable if the eigenvalue problem
produces purely imaginary $\lambda $. On the contrary, the existence of
eigenvalues with nonzero real parts implies instability. In the latter case,
it is of particular interest to simulate nonlinear evolution of unstable
states.

\subsection{The stationary variational approximation}

\label{sec:VA}

Before turning to numerical computations, it is natural to apply a
variational approximation (VA)~\cite{progopt} to the stationary equation (%
\ref{eq:st}). This will allow us to examine how well numerical solutions,
that are produced in the next section, can be approximated by the simple
(exponentially decaying) analytical expressions. To this end, we define%
\begin{equation}
\phi _{n}\equiv C^{1/\left( 2\sigma \right) }\chi _{n}~,~~\frac{\omega }{C}%
\equiv \Omega ,  \label{phichi}
\end{equation}%
and thus rewrite Eq. (\ref{eq:st}) as

\begin{equation}
(\chi _{n+1}+\chi _{n-1})-\frac{\left( 2+\Omega \right) \chi _{n}}{1+\chi
_{n}^{2\sigma }}=0,  \label{fraction}
\end{equation}%
which may generate bright solitons for $\Omega >0$.

The Lagrangian from which Eq. (\ref{fraction}) can be derived is%
\begin{equation}
L=\sum_{n=-\infty }^{+\infty }\left[ \chi _{n+1}\chi _{n}-\left( 2+\Omega
\right) \int \frac{\chi _{n}d\chi _{n}}{1+\chi _{n}^{2\sigma }}\right] ~.
\label{L}
\end{equation}%
The integral in this expression can be formally expressed in terms of the
hypergeometric function, similar to Eq. (\ref{eq:norm}), but such a formula
is not really needed.

The simplest ansatz for the stationary soliton follows the pattern of Ref.
\cite{malomed}:%
\begin{equation}
\chi _{n}=A\exp \left( -a|n|\right) ,  \label{ansatz}
\end{equation}%
with $a>0$. Then, the first term in Eq. (\ref{L}) is easily calculated,
following the substitution of ansatz (\ref{ansatz}):%
\begin{equation}
\sum_{n=-\infty }^{+\infty }\chi _{n+1}\chi _{n}=\frac{A^{2}}{\sinh a}.
\label{chichi}
\end{equation}%
The variational (Euler-Lagrange) equations intended to provide stationary
discrete solitons solutions, derived from Lagrangian (\ref{L}), in which
ansatz (\ref{ansatz}) is substituted, read:%
\begin{equation}
\frac{\partial L}{\partial A}=\frac{\partial L}{\partial a}=0.  \label{d/d}
\end{equation}%
The explicit form of the resulting equations is:%
\begin{eqnarray}
\frac{2}{\sinh a}-\left( 2+\Omega \right) \sum_{n=-\infty }^{+\infty }\frac{%
\exp \left( -2a|n|\right) }{1+A^{2\sigma }\exp \left( -2\sigma a|n|\right) }
&=&0,  \label{A} \\
\frac{\cosh a}{\sinh ^{2}a}-\left( 2+\Omega \right) \sum_{n=-\infty
}^{+\infty }\frac{|n|\exp \left( -2a|n|\right) }{1+A^{2\sigma }\exp \left(
-2\sigma a|n|\right) } &=&0.  \label{a}
\end{eqnarray}%
The system of Eqs. (\ref{A}) and (\ref{a}) can be solved numerically for variable $A$
and $a$ if $\Omega >0$ and $\sigma $ are given. Although the solution of this
system is still performed numerically, hence our approach is not fully
analytical, the VA provides a significant insight into the problem, as it
converts the infinite-dimensional original system into a much simpler system
of two equations (\ref{A}) and (\ref{a}) for $A$ and $a$.

\begin{figure}[tbp]
\begin{tabular}{ccc}
\includegraphics[width=6cm]{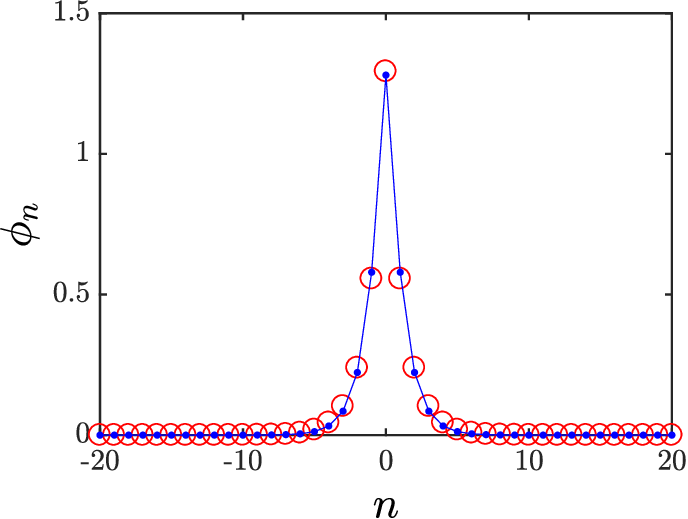} & %
\includegraphics[width=6cm]{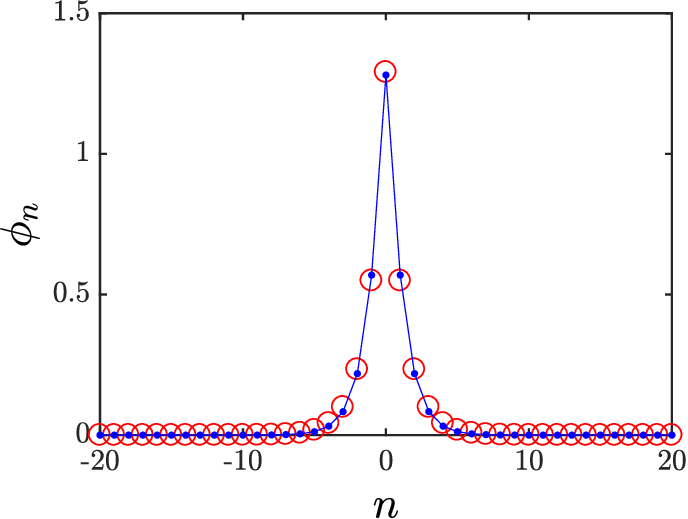} & %
\includegraphics[width=6cm]{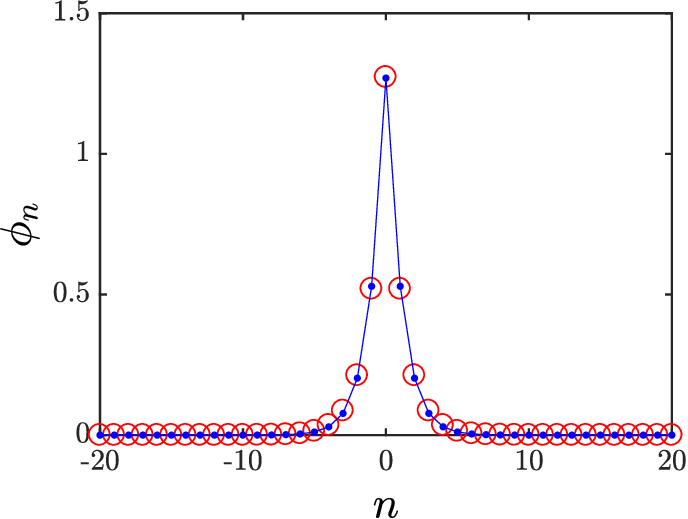} \\
\includegraphics[width=6cm]{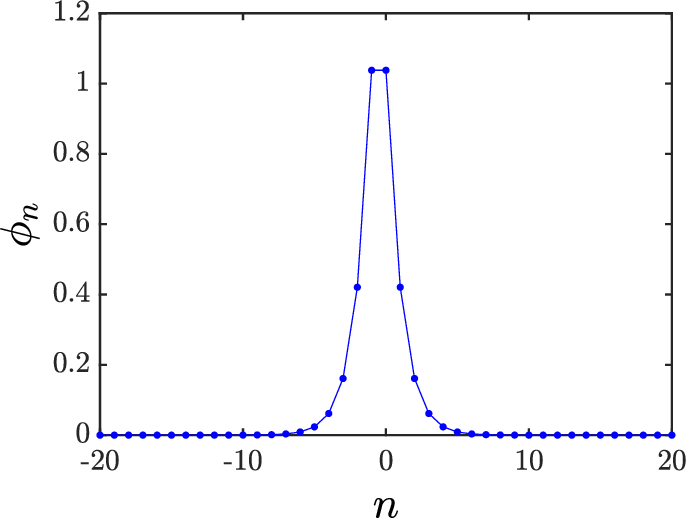} & %
\includegraphics[width=6cm]{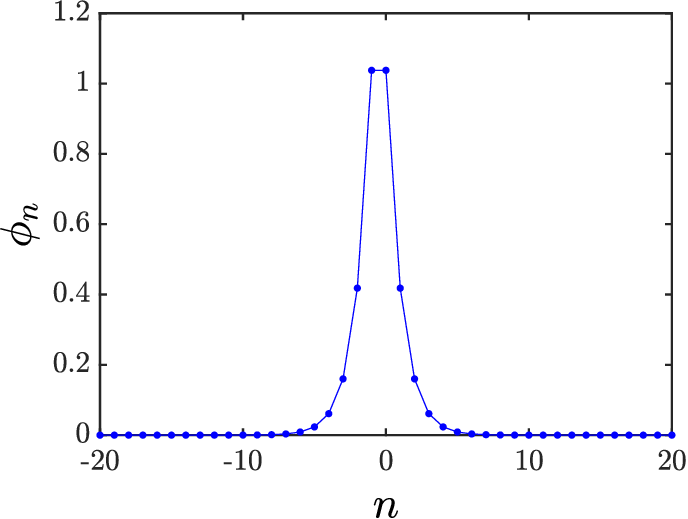} & %
\includegraphics[width=6cm]{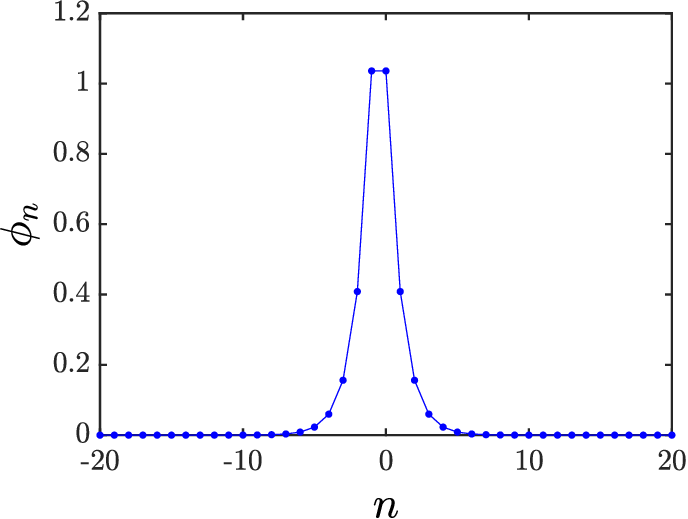} \\
\end{tabular}%
\caption{(Top panels) Exact numerically demonstrated profiles of the discrete
soliton (blue lines and dots), and their counterparts predicted by the
variational approximation (red circles), for 1-site solutions with $\protect%
\omega =1$ and $C=1$. The displayed solutions are characterized by the
respective values of $\protect\sigma $, namely $\protect\sigma =1.7$ (left),
$\protect\sigma =1.75$ (center) and $\protect\sigma =2$ (right). (Bottom panels) Numerical profiles of discrete 2-site solitons with the same parameters of the 1-site soliton above.}
\label{fig:profiles}
\end{figure}

\section{Numerical Results}

First of all, we consider the dependence of the stability of 1-site
(site-centered) and 2-site (inter-site centered) discrete solitons on the
nonlinearity power $\sigma $ for fixed frequency $\omega =1$ and coupling $%
C=1$. Such solutions, e.g., for both kinds of modes at $\sigma =1.7$, $%
\sigma =1.75$ and $\sigma =2$ are depicted in Fig.~\ref{fig:profiles}. 1-site solitons are compared therein with profiles predicted by the VA, demonstrating close agreement between the two; while the VA is formulated for 1-site solitons, the numerical profiles of the inter-site centered counterparts are also given for completeness.

\begin{figure}[tbp]
\begin{tabular}{ccc}
\includegraphics[width=6cm]{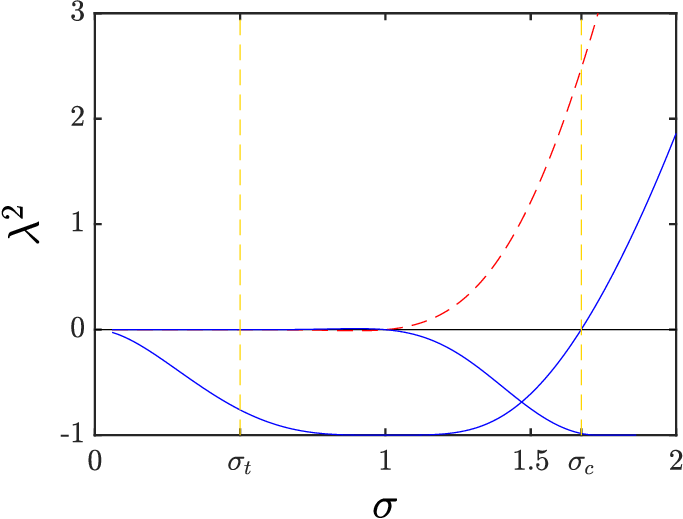} &
\includegraphics[width=6cm]{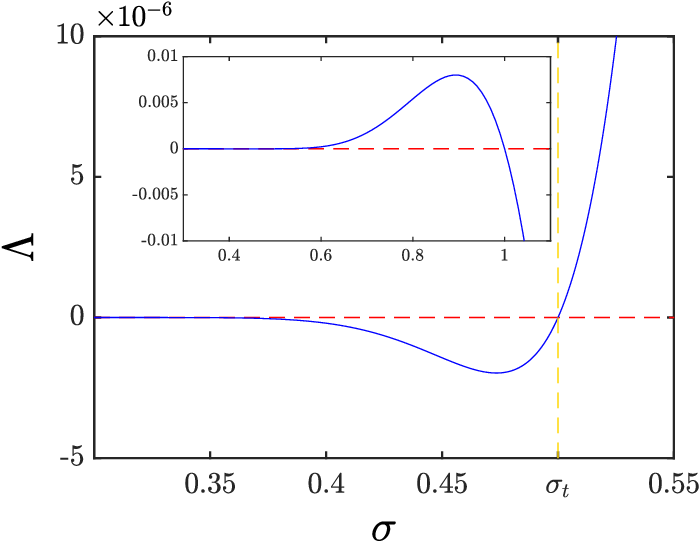}
\\ \vspace{0.5cm}
\includegraphics[width=6cm]{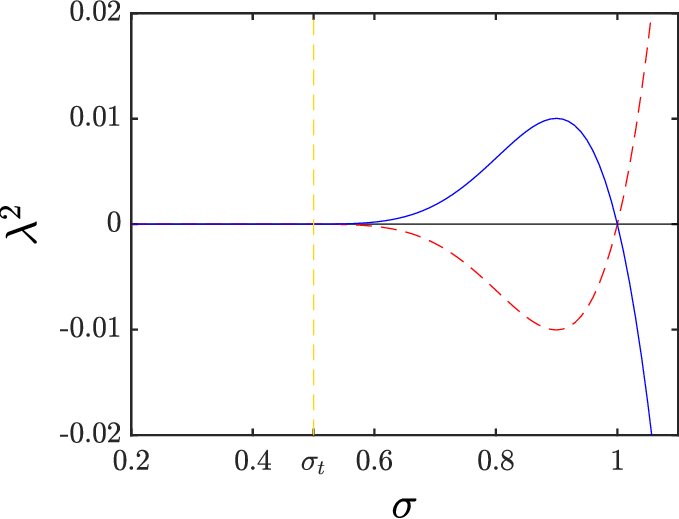} & %
\includegraphics[width=6cm]{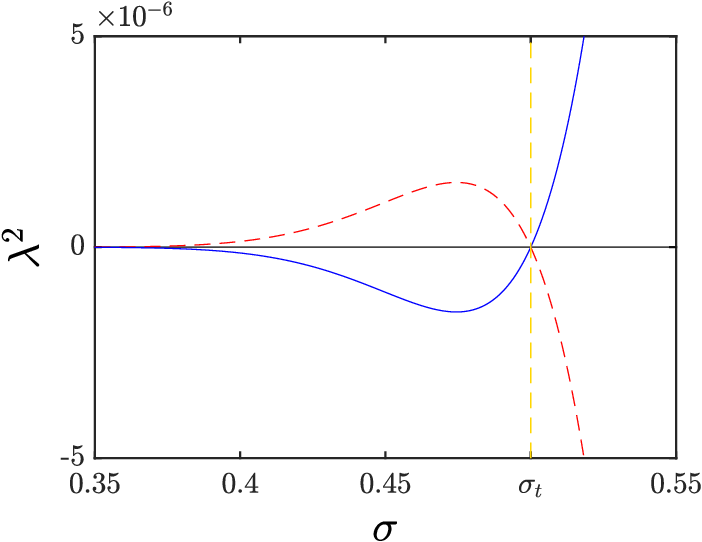} &  \\
&  &
\end{tabular}%
\caption{Top left panel: linear stability eigenvalues for 1-site (full blue line)
and 2-site (dashed red line) discrete solitons as a function of $\protect%
\sigma $ for $\protect\omega =C=1$; top right panel:
the dependence of the Hessian eigenvalue $\Lambda$ closest to zero for 1-site solitons with the same parameters set. Bottom panels provide zoom around points
$\protect\sigma =1$ (left panel) and $\protect\sigma =0.5$ (right panel) of
the top left panel, illustrating two stability exchanges occurring between the 1-
and 2-site modes in the\ present model. The critical
values of $\protect\sigma$ for which the main bifurcations take place,
namely $\protect\sigma_c=1.674$ and $\protect\sigma_t=0.5$ are highlighted
by means of vertical dashed lines.}
\label{fig:stab}
\end{figure}

As it is well known~\cite{book}, in the case of
the cubic ($\sigma =1$) AL equation, there are two eigenvalue pairs at zero,
as a consequence of the translational and phase (gauge) invariance of the
model. One of these invariances, namely the effective translational
invariance, is broken in the non-integrable case, $\sigma \neq 1$. We start
by analyzing the situation for $\sigma >1$. For 1-site solitons,
computations demonstrate that the eigenvalue corresponding to the
translational mode moves along the imaginary axis, whereas, at the same
time, another mode departs from the linear-mode band; at $\sigma =\sigma
_{c} $ (with $\sigma _{c}=1.674$ for the particular value of $C=1$), the
latter mode becomes associated with a real eigenvalue pair, hence the 1-site
solitons are unstable at $\sigma >\sigma _{c}$. For 2-site solitons, the
translational mode moves along the real axis immediately beyond the
integrable limit, consequently the relevant solutions are unstable for every
$\sigma >1$. The scenario is quite different for $\sigma <1$, where the
situation is reversed for 1-site and 2-site modes. In particular, in that
case, the translational mode moves along the real (imaginary) axis for the
1-site (2-site) soliton, giving rise to exchange of the stability between
them. It is worthy to note that a second stability exchange occurs at $%
\sigma =\sigma _{t}$, with $\sigma _{t}=0.5$ for \emph{all} values of $C$
(recall that $\sigma =0.5$ corresponds to the LHY nonlinearity in the
two-component BEC \cite{Grisha1}). All these phenomena are summarized in
Fig.~\ref{fig:stab}.

\begin{figure}[tbp]
\includegraphics[width=6cm]{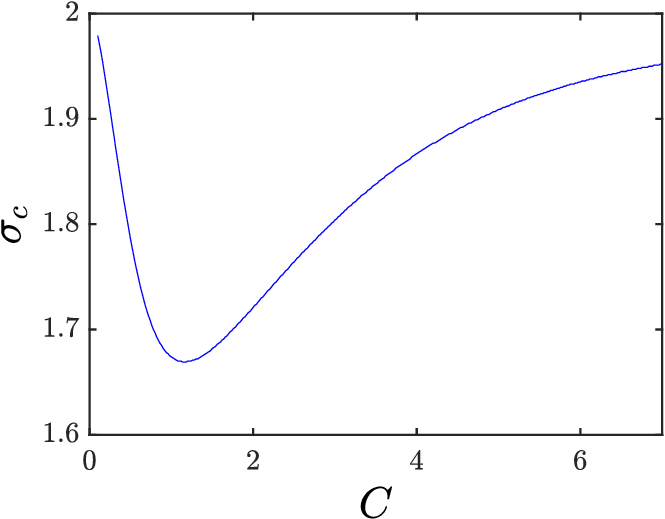}
\caption{The dependence of critical value $\protect\sigma _{c}$ on $C$ for $%
\protect\omega =1$.}
\label{fig:plane}
\end{figure}

\begin{figure}[tbp]
\begin{tabular}{cc}
\includegraphics[width=6cm]{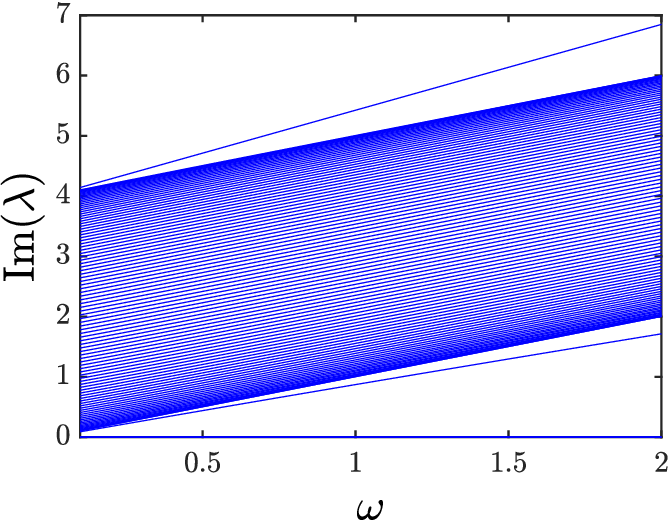} &
\includegraphics[width=6cm]{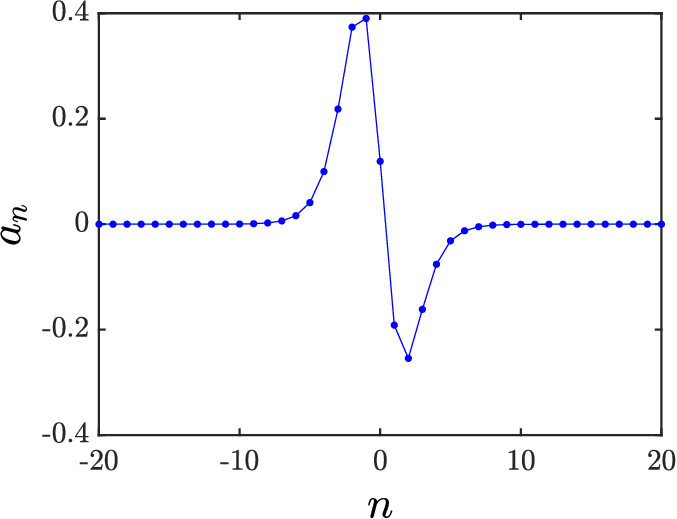}
\end{tabular}
\caption{(Left panel) The dependence of the imaginary part of the
stability eigenvalues for both 1-site and 2-site solitons with $C=1$ and $%
\protect\sigma =\protect\sigma _{t}=0.5$ with respect to $\protect\omega $. (Right panel) 0-eigenmode for the 1-site soliton with $\sigma=0.5$ and $C=\omega=1$}
\label{fig:sigmat}
\end{figure}

Secondly, we have analyzed the effect of varying the coupling constant on
the critical values $\sigma _{c}$ and $\sigma _{t}$. The approach helps to
reach the continuum limit, $C\rightarrow \infty $. Figure~\ref{fig:plane}
shows the dependence of $\sigma _{c}$ on $C$ for $\omega =1$. One can
observe that the minimum value of $\sigma _{c}$ is $\approx 1.669$, a value
that is attained at $C\approx 1.16$; furthermore, $\sigma _{c}$ tends to $2$
when $C\rightarrow 0$ and $C\rightarrow \infty $. The latter retrieves the
well-known continuum limit of the corresponding eigenvalue bifurcation
towards the collapse~\cite{sulem}. As mentioned in the above paragraph, the
critical value $\sigma _{t}$ is independent of $C$ and is equal to $1/2$.
Interestingly, the spectrum of solitons for $\sigma=\sigma_t=0.5$ is the same as in the integrable Ablowitz-Ladik lattice
$\sigma=1$ except for the occurrence of two (pairs of) localized eigenmodes,
one at the top and another one at the bottom of the linear modes band. Notice
that \emph{both} 1-site and 2-site solitons possess the same spectrum,as illustrated in Fig.~\ref{fig:sigmat}, and an associated neutral mode, as confirmed in Fig.~\ref{fig:stab}. Notice, however, that
the system in this case is not translationally invariant as the eigenvector that corresponds to the neutral mode (see right panel of the same figure) is not fully anti-symmetric, as is the case in the cubic ($\sigma=1$) limit.

\begin{figure}[tbp]
\begin{tabular}{cc}
\includegraphics[width=6cm]{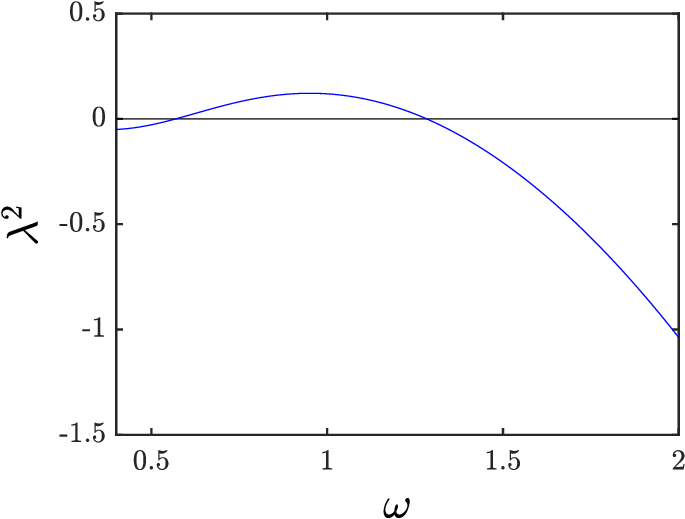} & %
\includegraphics[width=6cm]{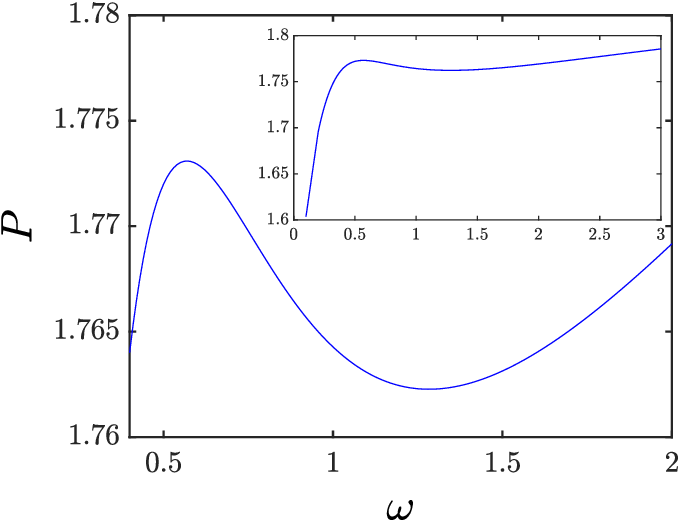} \\
&
\end{tabular}%
\caption{Dependences of the square of the relevant stability eigenvalue, $%
\protect\lambda^2$, and conserved quantity $P$ on frequency $\protect\omega$ (left and
right panels, respectively) for 1-site solitons with $C=1$ and $\protect%
\sigma =1.7$. The soliton is unstable in the interval of $\protect\omega \in
(0.56,1.29)$. Inset in right panel depicts the range $\protect\omega\in(0,3]$ confirming the existence of only two extrema in the $%
P(\protect\omega)$ curve.}
\label{fig:VK}
\end{figure}

Next we test the validity of the VK criterion in the present setting (see,
e.g.,~\cite{book} for its application to DNLS equations). It suggests that
the change of monotonicity of the frequency dependence of the
norm-like conserved quantity, defined
as per Eq.~(\ref{eq:norm}) amounts to a change of stability. To examine
this, we take 1-site solitons and vary $\omega $, fixing, e.g., $\sigma =1.7$
and $C=1$. The result is shown in Fig.~\ref{fig:VK}, where one can indeed
see the direct correlation between the slope of $P(\omega )$ dependence and
the actual stability, as expected from the VK criterion, and from the
general Grillakis-Shatah-Straus~theory \cite{grillakis}; see Ref. \cite{kapprom} for a recent overview. More specifically, the discrete solitons
are found to be stable (unstable) when the slope is positive (negative).
Additionally, in the case in which the emergence of an unstable linearization eigenvalue is not associated with a change of slope of the $P(\omega)$ curve one can compute the spectrum of the Hessian of the Hamiltonian (\ref{eq:Ham}) which can predict the existence of bifurcations through an eigenvalue zero crossing; this fact was illustrated in Fig.~\ref{fig:stab}.

\begin{figure}[tbp]
\begin{tabular}{ccc}
\includegraphics[width=6cm]{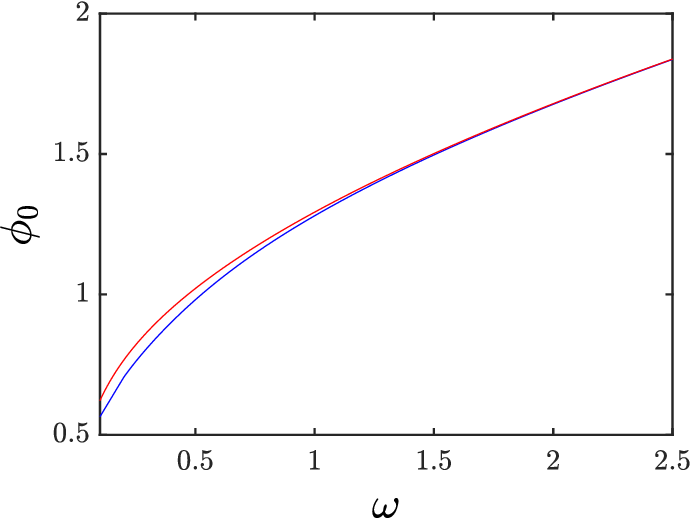} & %
\includegraphics[width=6cm]{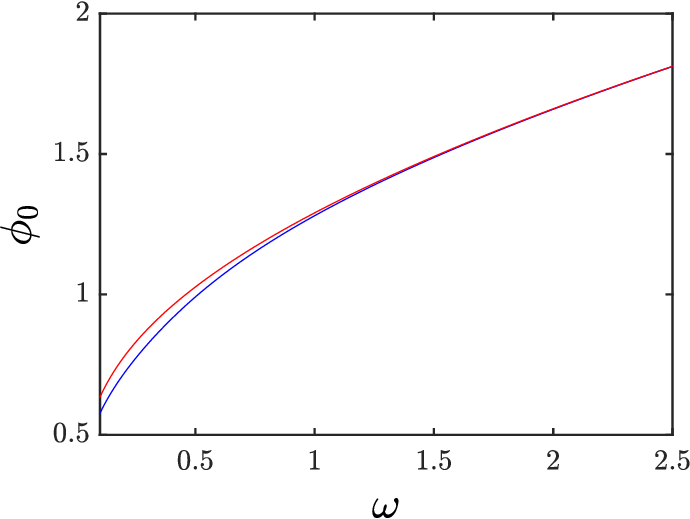} & %
\includegraphics[width=6cm]{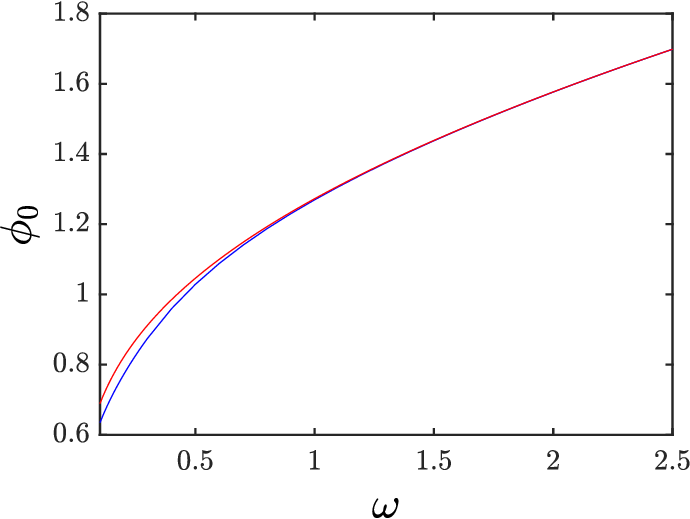} \\
\includegraphics[width=6cm]{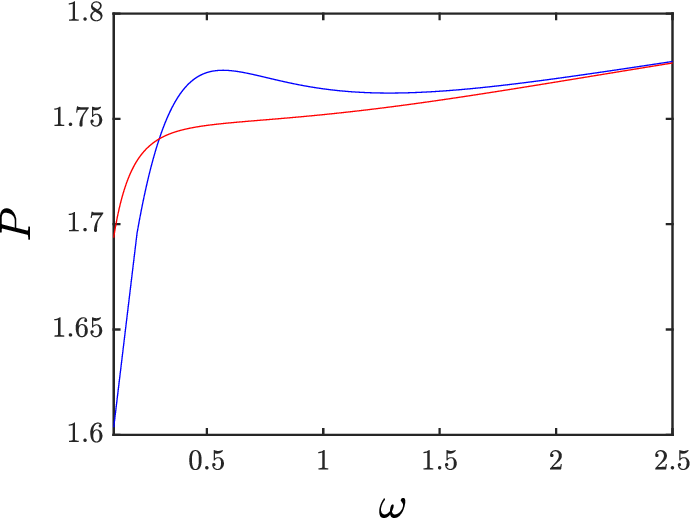} & %
\includegraphics[width=6cm]{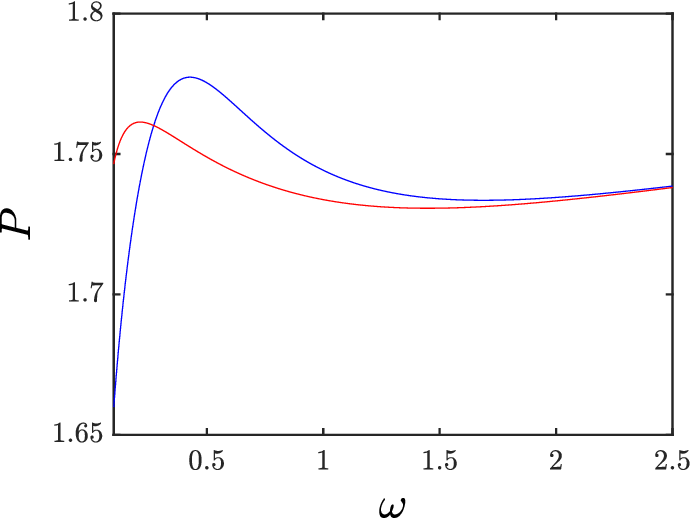} & %
\includegraphics[width=6cm]{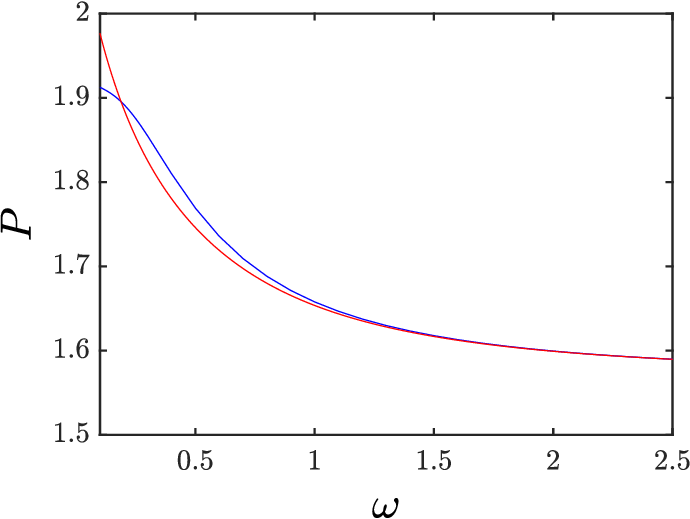} \\
&  &
\end{tabular}%
\caption{The amplitude ($\protect\phi _{0}$) and norm-like quantity
  $P$ of 1-site discrete
solitons (top and bottom panels, respectively) versus $\protect\omega $ for $%
\protect\sigma =1.7$ (left panels), $\protect\sigma =1.75$ (middle panels)
and $\protect\sigma =2$ (right panels), fixing $C=1$. Blue (red) line
corresponds to the numerical (variational) solution.}
\label{fig:VA}
\end{figure}

We have also compared the predictions of the VA, presented in subsection \ref%
{sec:VA}, with the results stemming from numerically exact (up to a
prescribed accuracy) calculations for the discrete solitons. Figure~\ref%
{fig:VA} shows the amplitude, $\phi_{0}$, and the conserved quantity $P$ of 1-site solitons
versus $\omega$, for three selected values of the nonlinearity power, $%
\sigma $ (namely, $\sigma =1.7$, $1.75$ and $2$) at fixed $C=1$; for other
values of $C$, the VA predictions can be obtained by the rescaling defined
in Eq. (\ref{phichi}). Note that, as seen in Fig.~\ref{fig:VA}, the VA
predicts quite accurately the amplitude of the discrete soliton in all the
cases considered, but there are some discrepancies in the dependence
of the conserved quantity $P$. In any case, the VA approach is more accurate for higher frequencies. In the limit
of $\omega \rightarrow \infty$ such models, including the DNLS equation, are tantamount to their so-called anti-continuum (AC),
isolated node limit. Here, the VA tends  to the corresponding exact  analytical solitary structure result, as shown in~\cite{Chong}.

More specifically, in the case of $\sigma =1.7$ the power is predicted to be
a monotonic function of $\omega $ by the VA, while the monotonicity is not
held by the numerical solution. This inaccuracy is a consequence of the
oversimplified ansatz adopted in Eq. (\ref{ansatz}); on the other hand, a
more sophisticated ansatz would not be amenable to the semi-analytical
treatment considered herein. As a result, the VA misses the bistability
interval and the associated changes in the stability. The situation is
better at $\sigma =1.75$, for which the VA correctly predicts the
nonmonotonicity, although the critical value of the frequency is not
accurately captured. Finally, for $\sigma =2$, both the VA and numerical
results yield a monotonic $P(\omega )$ dependence, in reasonable agreement
with each other. There is also some discrepancy in the value of $\sigma _{c}$%
, for which the VA predicts $\sigma _{c}=1.710$ with independence of the value of the
coupling constant, $C$.

\begin{figure}[tbp]
\begin{tabular}{cc}
\includegraphics[width=6cm]{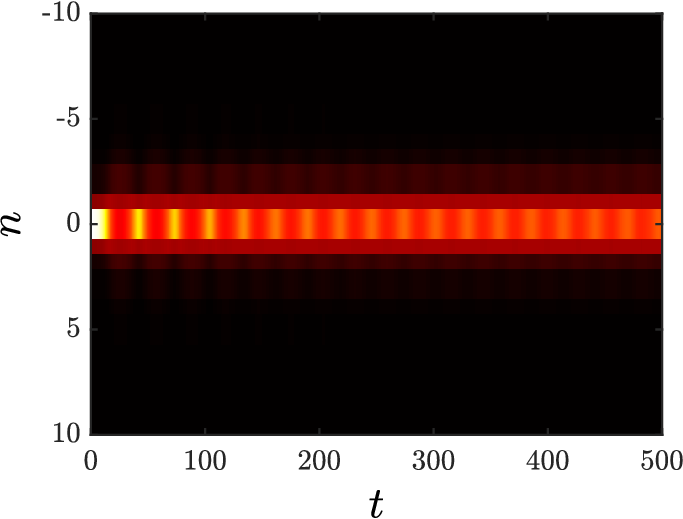} & %
\includegraphics[width=6cm]{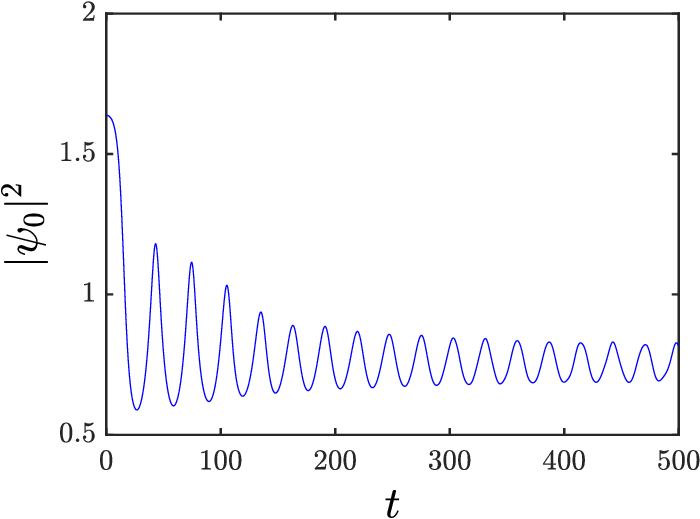} \vspace{.5cm} \\
\includegraphics[width=6cm]{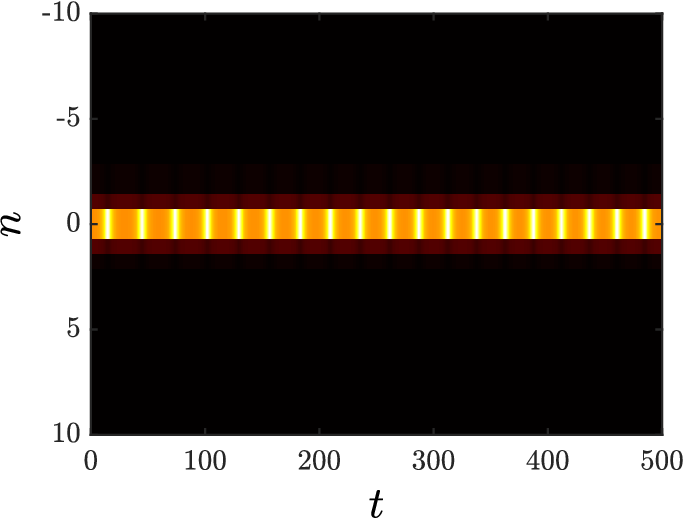} & %
\includegraphics[width=6cm]{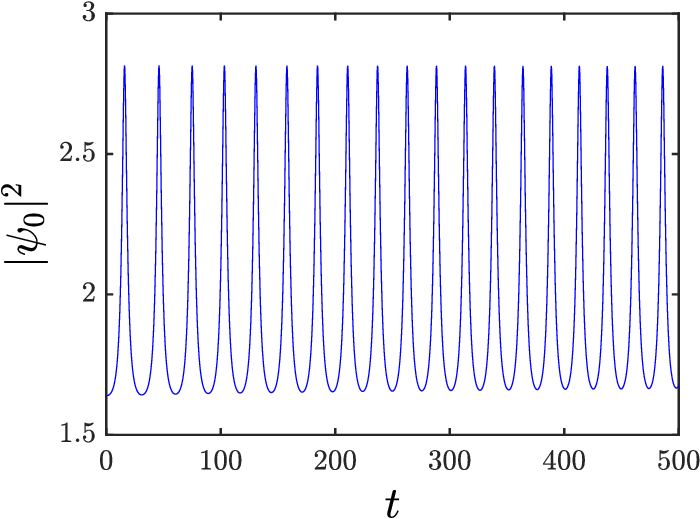} \vspace{.5cm} \\
&
\end{tabular}%
\caption{{The evolution of the soliton density (left panel) and central-site
density (right panel) for 1-site solitons with $\protect\omega =C=1$ and $%
\protect\sigma =1.7$. In the top (bottom) row, the soliton is perturbed
along direction $\{a_{n}\}$ ($\{b_{n}\}$), with $[\{a_{n}\},\{b_{n}\}]$
being the eigenvector corresponding to the most unstable eigenvalue.}}
\label{fig:dyn1a}
\end{figure}

\begin{figure}[tbp]
\begin{tabular}{cc}
\includegraphics[width=6cm]{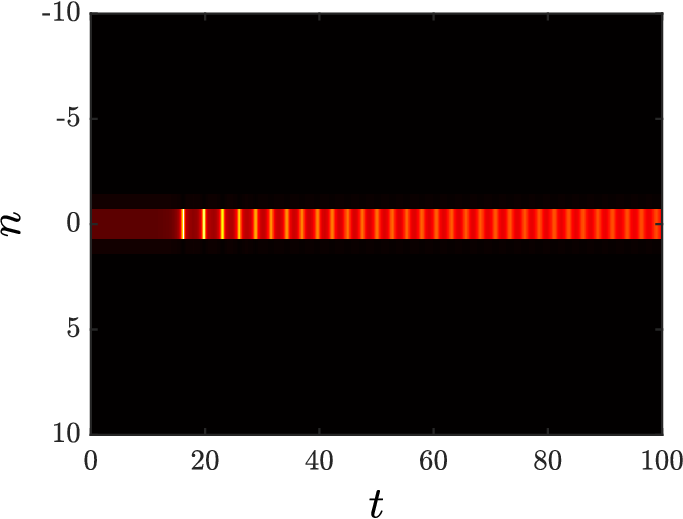} & %
\includegraphics[width=6cm]{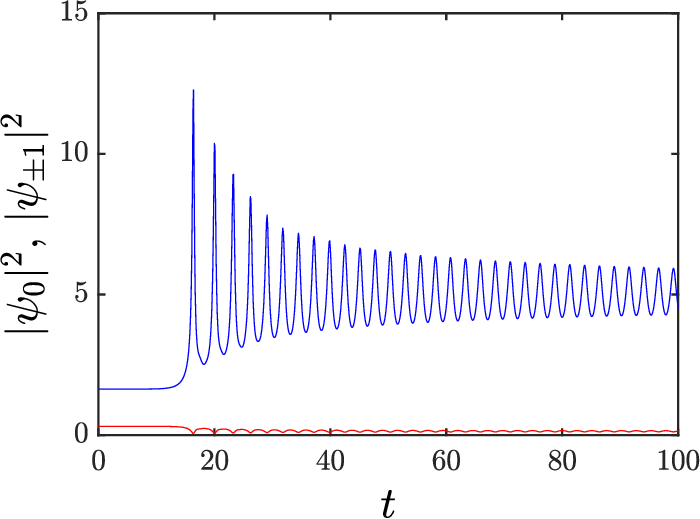} \vspace{.5cm} \\
\end{tabular}%
\caption{The evolution of the soliton density (left panel) and the central
sites density (right panel) for the 1-site soliton with $\protect\omega =C=1$
and $\protect\sigma =1.8$. A
small random perturbation was added to the (unstable) discrete solitons.}
\label{fig:dyn1b}
\end{figure}

Finally, we mention the main features of unstable discrete
solitons evolution for different typical examples of the instability that are
identified above. Figures~\ref{fig:dyn1a} and \ref{fig:dyn1b} show the outcome
of the evolution for unstable 1-site solitons with $\sigma >1$: it is
observed that the soliton density at $n=0$, $|\psi _{0}|^{2}$, deviates from
the equilibrium value and subsequently performs oscillations around a new
equilibrium value. In cases when bistability takes place, as, e.g., at $\sigma
=1.7$ in Fig.~\ref{fig:dyn1a}, the unstable dynamical evolution, depending
on a particular choice of the initial perturbation, may result in
oscillations around states belonging, in the top and bottom panels, to two
different branches of stable solutions. One can also observe that the soliton sustains together with the increasing of the density at $n=0$,
a narrowing during the oscillations that manifest e.g. as a simultaneous decreasing of $|\psi_{\pm1}|^2$. In addition, the maximum of $|\psi _{0}|^{2}$
increases with $\sigma $, and for $\sigma=1.8634$ it leads to computational overflow. Beyond this limit, despite using multiple algorithms,
we have been unable to converge to a solution. Because of this, we cannot go beyond this value of $\sigma$ in our simulations. Our phenomenology basically involves an extremely fast increasing of $|\psi_0|^2$ that cannot be compensated by the decreasing of the term $(\psi_1+\psi_{-1})$ and hence, our
numerical integrators fail to converge. It is important to note here that {\it contrary} to the DNLS case or the cubic nonlinearity limit (i.e. of logarithmic divergence), the conservation laws in the present case do not a priori preclude the density of a single site from becoming infinite. Clearly, identifying the phenomenology of the model for sufficiently large $\sigma$ is an important open topic that merits considerable additional study including from a theoretical point of view.

\begin{figure}[tbp]
\begin{tabular}{cc}
\includegraphics[width=6cm]{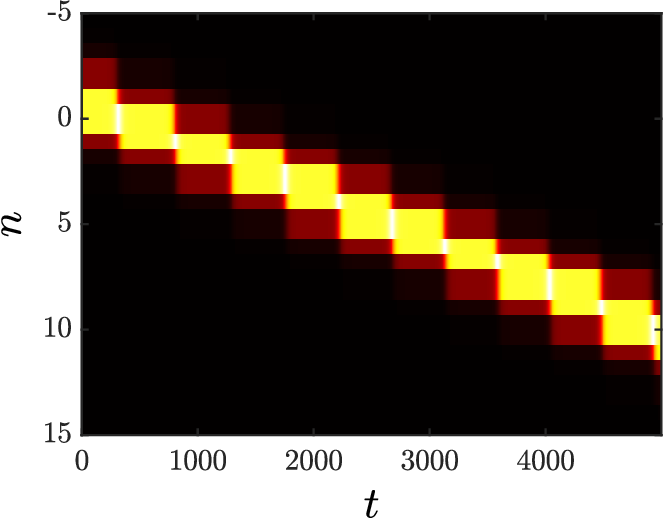} & %
\includegraphics[width=6cm]{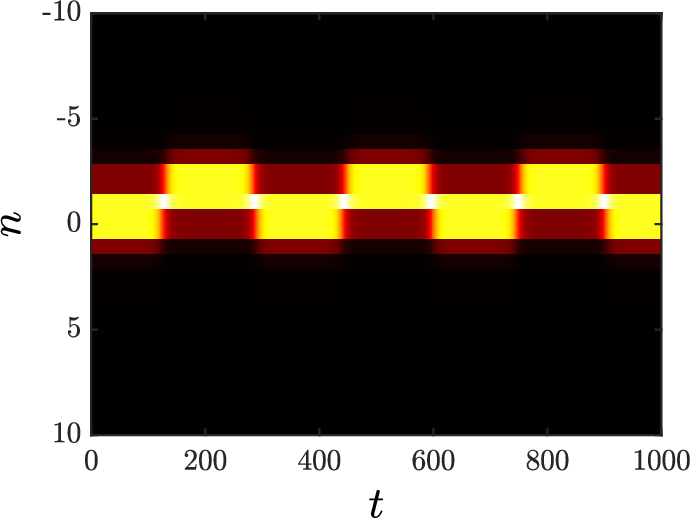} \vspace{.5cm} \\
&
\end{tabular}%
\caption{The evolution of the soliton density for 2-site solitons with $\protect\omega =C=1$
and $\protect\sigma =1.01$ (left panel), and $\protect\sigma =1.05$ (right panel). In this case too, a small
random perturbation was added to the unstable discrete solitons.}
\label{fig:dyn2a}
\end{figure}

\begin{figure}[tbp]
\begin{tabular}{cc}
\includegraphics[width=6cm]{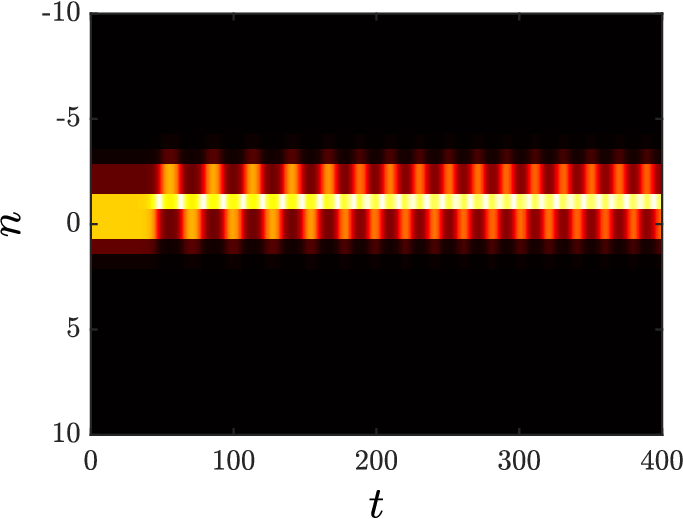} & %
\includegraphics[width=6cm]{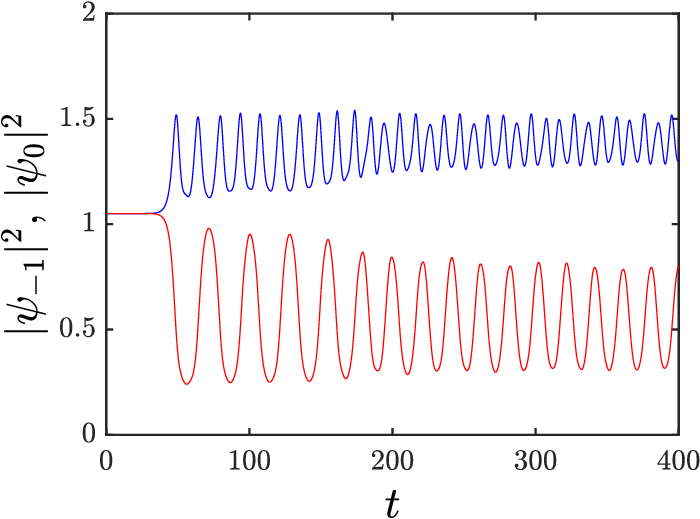} \vspace{.5cm} \\
\includegraphics[width=6cm]{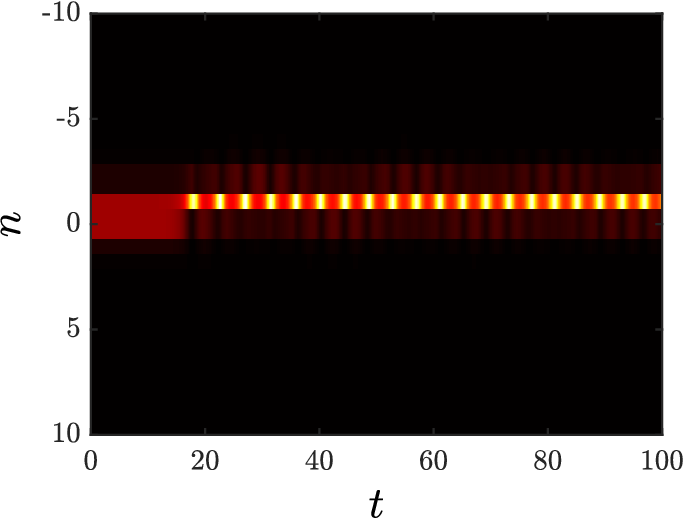} & %
\includegraphics[width=6cm]{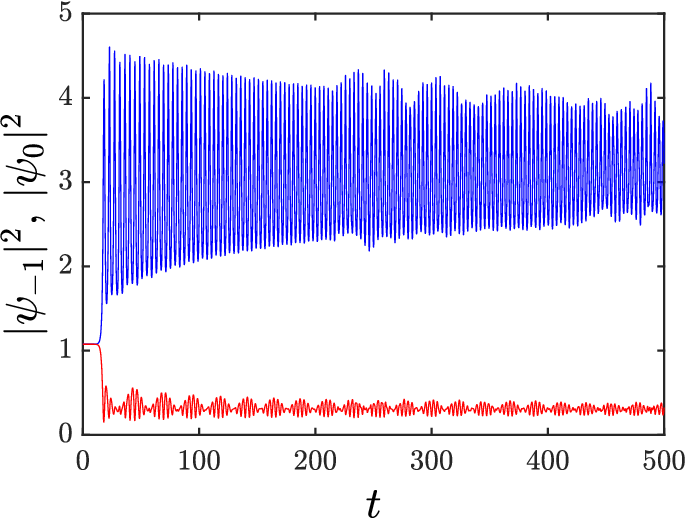} \vspace{.5cm} \\
\includegraphics[width=6cm]{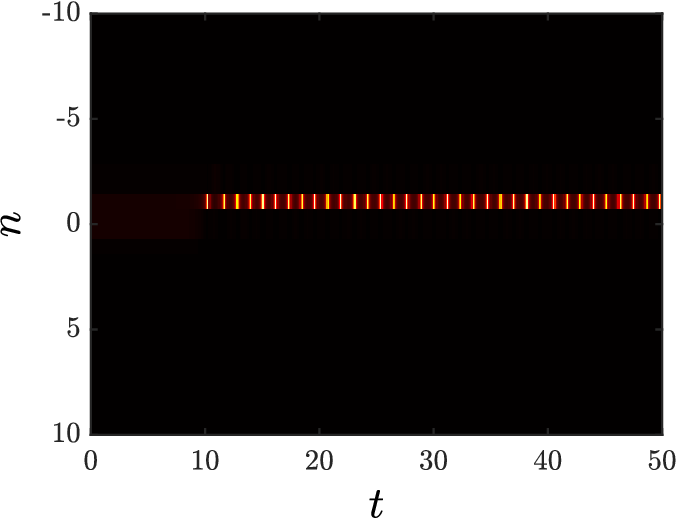} & %
\includegraphics[width=6cm]{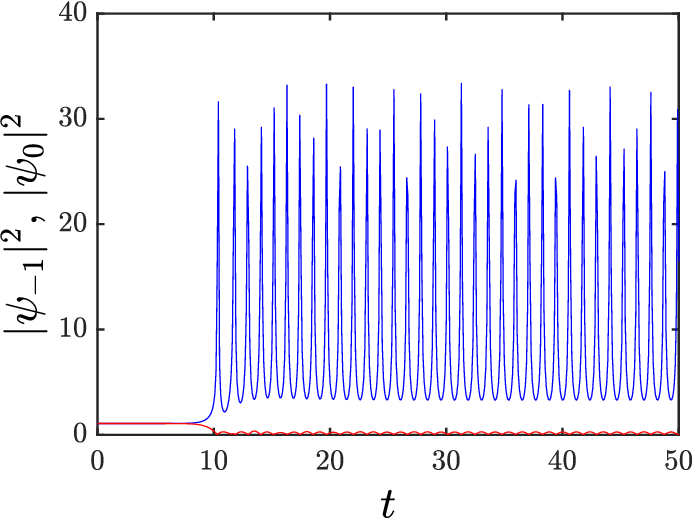} \vspace{.5cm} \\
&
\end{tabular}%
\caption{The evolution of the soliton density (left panel) and the central
sites density (right panel) for 2-site solitons with $\protect\omega =C=1$
and $\protect\sigma =1.2$, $\protect\sigma =1.5$, or $\protect\sigma =1.7$
(top, middle, and bottom rows, respectively). In this case too, a small
random perturbation was added to {the unstable discrete solitons}.}
\label{fig:dyn2}
\end{figure}

The dynamics of unstable 2-site solitons with $\sigma >1$ is somewhat different when $\sigma$ is close to 1. If $\sigma$ is close enough to 1 (as e.g. $\sigma=1.01$ for $\omega=C=1$), the soliton becomes mobile but, if $\sigma-1$ is high enough (as e.g. $\sigma=1.05$ for $\omega=C=1$), the soliton gets pinned and its profile oscillates between a 1-site and 2-site soliton (see Fig.~\ref{fig:dyn2a}). When $\sigma$ is increased (as e.g. to $\sigma=1.2$ for $\omega=C=1$), one can observe that the soliton undergoes a breaking of the symmetry between its two central sites (here, these are $n=-1$ and $n=0$). As a
result, a dominant fraction of the mass concentrates at one of the two sites, e.g., at $n=0$, although a relatively high density remains at $n=-1$
too, decreasing with the increase of $\sigma $ (see Fig.~\ref{fig:dyn2}). The high density fraction that concentrates at $n=0$ experiences a similar trend to the observed in the 1-site case.

It is also worthy to mention that unstable 1-site solitons with $\sigma <1$
feature mobility in the underlying lattice (not shown here in detail), which
is a consequence of the stability-exchange bifurcation associated with the
translational perturbation mode.

\section{Conclusions and future challenges}

In the present work we have explored a non-integrable generalization of the
AL (Ablowitz-Ladik) model, bearing the power-law nonlinearity. The model by
itself is interesting for a variety of reasons, such as the discretization
of the continuum problem with the power-law nonlinearity, an extension of
the integrable AL model, the mean-field limit of the Bose-Hubbard lattice
with the population-dependent inter-site coupling, and as a testbed for
addressing the interplay of discreteness and the potential for
collapse (for a sufficiently
large nonlinearity power, $\sigma $).

Our analysis has revealed essential features of the model. We have
identified its conservation laws, including an unprecedented form of the
conserved quantity (which falls back to the well-known logarithmic expression for the AL
norm at $\sigma =1$) and the Hamiltonian. We have demonstrated how the
modified norm-like quantity can be used (through its frequency dependence) to predict
potential changes of stability, by means of the extended VK
(Vakhitov-Kolokolov) criterion. We have also formulated the VA\ (variational
approximation) based on the exponential ansatz, which is well capable of
capturing the amplitude of the discrete solitons, but is considerably less
accurate (due to limitations imposed by the fixed form of the analytically
tractable ansatz) in capturing the quantity $P$, and, hence, the corresponding
VK-predicted stability changes. We have identified a number of stability
alternations between the 1- and 2-site (alias site- and inter-site-centered)
modes, and addressed their dynamical implications. Importantly, for
sufficiently large $\sigma $, the present model, on the contrary to the
standard DNLS equation,
appears to feature extreme amplitudes, which are not precluded
  a priori from the conservation laws. Whether indeed some form
  of collapse is present is an intriguing question that merits
  further study both from a theoretical and from a numerical perspective.

The present study paves the way for numerous questions worthwhile of further
examination. Elaborating a better variational ansatz, if it may be (semi-)
analytically tractable, will certainly help to develop more accurate
understanding of the model's features. An exploration of stability
switchings, such as the one at the above-mentioned stability-exchange point $%
\sigma _{t}$, and of the origin of these phenomena, are certainly relevant
issues in their own right. Extending the model to higher dimensions, and, in
particular, the study of phenomenology of discrete vortices~\cite{book} is
relevant too, while higher-dimensional AL models were, thus far, addressed
in a very limited set of studies~\cite{lafortune,xiaowu}. Lastly, as regards
the focusing case with $\beta =1$, it should be interesting to address a
possibility of the existence of rogue-wave modes, such as the (discrete
variant of the) Peregrine soliton~\cite{akhm,yang}. At the same time,
exploring the defocusing case with $\beta =-1$ and the corresponding
discrete dark solitons, as well as delocalized vortices, would be a subject
of interest in its own right. Efforts along some of these directions are
currently in progress and will be reported in future studies.

\vspace{2mm}

\textit{Acknowledgements.} J.C.-M.~thanks the financial support from
MAT2016-79866-R project (AEI/FEDER, UE). P.G.K. acknowledges the support by
NPRP grant \# [9-329-1-067] from Qatar National Research Fund (a member of
Qatar Foundation). The work of B.A.M. was supported, in part, by the Israel
Science Foundation, through grant No. 1287/17. This material is also based
upon work supported by the National Science Foundation under Grant No.
PHY-1602994 (P.G.K.).

Findings reported herein are solely the responsibility of the authors.


\begin{thebibliography}{99}
\bibitem{book} P. G. Kevrekidis, \textit{The Discrete Nonlinear Schr{\"{o}}%
dinger Equation}, Springer-Verlag (Heidelberg, 2009).

\bibitem{sulem} C. Sulem and P. L. Sulem,
\newblock  {\it The Nonlinear
Schr{\"o}dinger Equation} (Springer-Verlag, New York, 1999).

\bibitem{ablowitz} M. J. Ablowitz, B. Prinari, and A. D. Trubatch, \textit{%
Discrete and Continuous Nonlinear Schr{\"{o}}dinger Systems}, Cambridge
University Press (Cambridge, 2004).

\bibitem{ussiam} P. G. Kevrekidis, D. J. Frantzeskakis, and R. Carretero-Gonz%
{\'a}lez, \textit{The defocusing Nonlinear Schr{\"o}dinger Equation: From
Dark Solitons to Vortices and Vortex Rings} (SIAM, Philadelphia, 2015).

\bibitem{dnc} D. N.\ Christodoulides, F.\ Lederer, and Y.\ Silberberg,
Nature \textbf{424}, 817 (2003); A. A.\ Sukhorukov, Y. S.\ Kivshar, H. S.\
Eisenberg, and Y.\ Silberberg, IEEE J. Quant. Elect. \textbf{39}, 31 (2003).

\bibitem{moti} F. Lederer, G. I. Stegeman, D. N. Christodoulides, G.
Assanto, M. Segev, and Y. Silberberg, Phys. Rep. \textbf{463}, 1 (2008).

\bibitem{ober} O. Morsch and M. Oberthaler, Rev. Mod. Phys. \textbf{78}, 179
(2006).

\bibitem{Peybi} M. Peyrard, Nonlinearity \textbf{17}, R1 (2004).

\bibitem{chong} C. Chong, P. G. Kevrekidis, \textit{Coherent Structures in
Granular Crystals: From Experiment and Modelling to Computation and
Mathematical Analysis}, Springer Verlag (Heidelberg, 2018).

\bibitem{niemi} A. K. Sieradzan, A. Niemi, and X. Peng, Phys. Rev. E \textbf{%
90}, 062717 (2004).

\bibitem{ablolad} M. J. Ablowitz, J. F. Ladik, J. Math. Phys. \textbf{16},
598 (1975); \textit{ibid.} \textbf{17}, 1011 (1976).

\bibitem{kapitula} T. Kapitula, P. Kevrekidis, Nonlinearity \textbf{14}, 533
(2001).

\bibitem{bishop} D. Cai, A. R. Bishop, N. Gr{\o }nbech-Jensen, and M.
Salerno Phys. Rev. Lett. \textbf{74}, 1186 (1995).

\bibitem{salerno} M. Salerno, Phys. Rev. A \textbf{46}, 6856 (1992).

\bibitem{cai} D. Cai, A. R. Bishop, and N. Gr{\o }nbech-Jensen Phys. Rev. E
\textbf{53}, 4131 (1996).

\bibitem{review} O. Dutta, M. Gajda, P. Hauke, M. Lewenstein, D.-S. Luhmann,
B. Malomed, T. Sowinski, and J. Zakrzewski, Rep. Prog. Phys. \textbf{78},
066001 (2015).

\bibitem{dmitriev} S. V. Dmitriev, P. G. Kevrekidis, B. A. Malomed, and D.
J. Frantzeskakis Phys. Rev. E \textbf{68}, 056603 (2003).

\bibitem{akhm} A. Ankiewicz, N. Akhmediev, and J. M. Soto-Crespo Phys. Rev.
E \textbf{82}, 026602 (2010).

\bibitem{yang} Y. Ohta and J. Yang, J. Phys. A: Math. Theor. \textbf{47},
255201 (2014).

\bibitem{hoffman} C. Hoffmann, E. G. Charalampidis, D. J. Frantzeskakis, P.
G. Kevrekidis, arXiv:1710.04899.

\bibitem{Turitsyn} E. W. Laedke, K. H. Spatschek, and S. K. Turitsyn, Phys.
Rev. Lett. \textbf{73}, 1055 (1994).

\bibitem{malomed} B.A. Malomed, M. I. Weinstein, Phys. Lett. A \textbf{220},
91 (1996).

\bibitem{weinstein} M. I. Weinstein, Nonlinearity \textbf{12}, 673 (1999).

\bibitem{ourDNLS} J. Cuevas, P. G. Kevrekidis, D. J. Frantzeskakis, B.A.
Malomed, Physica D \textbf{238}, 67 (2009).

\bibitem{Kladko} S. Flach, K. Kladko, R. S. MacKay, Phys. Rev. Lett. \textbf{%
78}, 1207 (1997).

\bibitem{Grisha1} D. S. Petrov and G. E. Astrakharchik, Phys. Rev. Lett.
\textbf{117}, 100401 (2016).

\bibitem{Grisha2} G. E. Astrakharchik and B. A. Malomed, Phys. Rev. A
\textbf{98}, 013631 (2018).

\bibitem{Vakh} M. Vakhitov and A. Kolokolov, Radiophys. Quantum Electron.
\textbf{16}, 783 (1973).

\bibitem{progopt} B.A. Malomed, Progr. Opt. \textbf{43}, 69 (2002).

\bibitem{grillakis} M. Grillakis, J. Shatah, W. Strauss, J. Funct. Anal.
\textbf{74}, 160 (1987); J. Funct. Anal. \textbf{94}, 308 (1990).

\bibitem{Chong} C. Chong, D. E. Pelinovsky, G. Schneider, Physica D \textbf{%
241}, 115 (2012).

\bibitem{kapprom} T. Kapitula, K. Promislow, \textit{Spectral and dynamical
stability of nonlinear waves}, Springer-Verlag (New York, 2013).

\bibitem{lafortune} P. G. Kevrekidis, G. J. Herring, S. Lafortune, Q. E.
Hoq, Phys. Lett. A \textbf{376}, 982 (2012).

\bibitem{xiaowu} X. Y. Wu, B. Tian, L. Liu, Y. Sun, Comm. Nonlin. Sci. Num.
Simul. \textbf{50}, 201 (2017).
\end{thebibliography}
\end{document}